\documentclass [12pt]{article}
\usepackage{lscape}                                           %
\usepackage[centertags]{amsmath}
\usepackage{amsfonts} \usepackage{amssymb}%\usepackage{eufrak}
\usepackage{color}
\usepackage{graphicx}
\usepackage{lscape}
\newcommand{\cl}{{l}}%{{\cal l}}
\newcommand{\T}{{T_0}}
\newcommand{\s}[1]{_{(#1)}}

\newcommand{\cP}{{\cal P}}

\newcommand{\Z}{\mathbb{Z}}

\newcommand{\oh}{\frac{1}{2}}

\newcommand{\COMMENTO}[1]{}
\newcommand{\COMMENTOO}[1]{}
\begin{document}
%%%%%%%%%%%%%%%%%%%%%%%%%%%%%%%%%%%%%%%%%%%%%%%%%%%%%%%%%%%%%%%%%%%%%%%%%
\title{
Light cone quantization and interactions of a new closed bosonic
string inspired to $D1$ string.
}

\author{%\parbox{11.5cm}
{Igor Pesando$^1$}
\\
~\\
~\\
$^1$Dipartimento di Fisica, Universit\`a di Torino\\
and I.N.F.N. - sezione di Torino \\
Via P. Giuria 1, I-10125 Torino, Italy\\
\vspace{0.3cm}
\\{ipesando@to.infn.it}
}

\maketitle
\thispagestyle{empty}

\abstract{
We quantize %a bosonic string with a worldvolume electric field
the bosonic part of the $D1$ string with closed boundary conditions
on the light cone and we consider
the $U(1)$ worldsheet gauge field a dynamical variable.
We compute also 3-Reggeon vertex by the overlapping technique.
We find that the Fock space is the sum of sectors
characterized by the momentum of the $U(1)$ Wilson line and that  
these sectors do not interact among them.
Each sector has exactly the same spectrum of the usual bosonic string 
when expressed in properly sector dependent rescaled variables. 
Rescaling is forced by factorization of the string amplitudes. 
We are also able to determine the relative string coupling constant 
of the different sectors.
It follows a somewhat  unexpected picture in which the effective action is
always the same independently on the sector 
but string amplitudes are only the same when expressed in sector
dependent rescaled variables. 
}
\\
keywords: {Bosonic string, D-brane}
%\\
%preprint:{DFTT--2013}

\newpage

\section{Introduction}
Since the discovery of branes and their role in the non perturbative
physics of string theory quite a lot of attention has been
devoted to understanding their dynamics and in particular 
building their actions.
Type IIB string enjoys a non perturbative $SL(2,\Z)$ invariance in 10 dimensions
therefore one of the aims was to construct D-brane actions 
(see for example \cite{Schwarz:1995dk}, \cite{sl2z},
\cite{Bergshoeff}) as well as to
compute low energy actions  which respect this symmetry
(see for example \cite{sl2zeffact}).
The need of checking these effective actions directly from string
computations has led to the development of the instanton calculus in
string theory (see for example \cite{inst}) on one side and 
to the study of the action of the S-duality on string
amplitudes on the other (see for example
\cite{Garousi0},\cite{Garousi}).
% and the formulation of ansatz\"e for the full $SL(2,\Z)$ invariant string amplitudes.

Among all branes in type IIB $D1(p,q)$ branes play a prominent role since
they are close relatives of the fundamental string and
therefore they are very relevant in
exploring the non perturbative picture of the IIB superstring.
One can then wonder whether $D1(p,q)$ brane actions can be taken as
fundamental actions and not only as effective ones.
There were studies of the quantization of the $D1$ action 
performed in \cite{Kallosh:1997nr} whose main interest was anyhow to
show that it is possible to quantize Green-Schwarz $D1$ string in a
covariant gauge. 
Nevertheless there were also studies more in line with the previous
question and they explored the quantization of the
$SL(2,\Z)$ invariant action for the $D1(p,q)$ \cite{Chalmers:1999ap}
in the hope of being able to recover the full  $SL(2,\Z)$ invariant
string amplitudes directly from prime  principles.

The aim of this paper is less ambitious and it is to consider the
bosonic part of the $D1$ string action as a fundamental action and to
quantize it as a first step to understand
what the full $\kappa$-symmetric $D1$ action can teach us.
Eventually assuming the action of the $D1$ brane as a fundamental
action and not an effective one  it could happen that studying its
interactions one could compute perturbatively in $1/R$ 
the strong coupled amplitudes in $AdS_5 \times S_5$ and
compare them directly with super Yang-Mills results.

In this paper 
we consider the action with closed string boundary conditions and we
quantize it on the light cone because the covariant quantization is more
difficult to implement than usual since the constraint algebra
involves structure functions and not only structure constants.
 
%More precisely we show that the covariant quantization
%while desiderable is trickier than usual because the constraint
%algebra involves structure functions and therefore we proceed with
%light cone quantization.
Then we compute the spectrum and the critical dimension which turn
out to be the same of the normal bosonic string.
Our result for the spectrum is at variance from those in the literature 
\cite{Kallosh:1997nr} since we find that it always contain  massless
states. We can track down the reason of this difference to the fact
that in those papers it was assumed that the worldsheet electric field
strength is diagonal in the Fock space while it is actually not.
We also find that the Fock space is the direct sum of Fock subspaces 
characterized by the momentum of the $U(1)$ Wilson line
 and that each Fock subspace describes
particles with the same mass spectrum as the usual bosonic string.

Finally we exam the interactions and find that 
each sector interacts only with its self and this
explains some negative results found in the literature (see for
example \cite{Kar:1998yv}) where it was found that it is not possible
to make $D$ strings and $F$ strings interact perturbatively.
We show that all sectors have the same interactions as the bosonic string 
up to a different string coupling constant whose relative
normalization we determine.

\section{The action and its quantization}
We consider the bosonic action
\begin{align}
S= -\T \int d^2 \xi 
\left[
\sqrt{ -\det (\hat G_{\alpha\beta} +\kappa  F_{\alpha\beta})}
+\theta F_{0 1}
\right]
\end{align}
where $\alpha,\beta,\dots\in\{0,1\}$ are worldsheet indexes,
$\hat G_{\alpha\beta}= G_{\mu \nu} \partial_\alpha X^\mu \partial_\beta X^\nu$ 
is the pull back of the spacetime metric, $\mu,\nu,\dots $ are
spacetime indexes
and  $F_{0 1}= \partial_0 A_1-\partial_1 A_0$ 
is the worldsheet $U(1)$ electric field.
$\kappa$ is a constant and $\theta$ is a kind of $\theta$-vacuum constant.
We impose  closed boundary conditions, i.e. 
$X(\sigma+\cl,\tau) = X(\sigma,\tau)$ and
$F_{01}(\sigma+\cl,\tau) = F_{01}(\sigma,\tau)$.

This action can be written in a Brink-Deser-Di Vecchia-Howe-Zumino \cite{Brink:1976sc} like form
with an additional worldsheet cosmological constant $\lambda$ as
\begin{align}
S= -\oh \T \int d\tau \int_0^\cl d \sigma
\left[
\sqrt{ -\det \gamma_{\alpha\beta} }~\gamma^{\alpha\beta}
(\hat G_{\alpha\beta} +\kappa  F_{\alpha\beta})
+\lambda \sqrt{ -\det g_{\alpha\beta}}
+2 \theta F_{0 1}
\right]
\end{align}
where $g_{\alpha \beta}$ is the worldsheet metric and
the matrix $\gamma_{\alpha \beta}=g_{\alpha \beta}+
\epsilon_{\alpha \beta} f $ with $f$  a scalar density 
is neither symmetric nor antisymmetric.
Because of this property the worldsheet supersymmetric action cannot be written
immediately.  

As usual at the classical level we must set $\lambda=0$ if we want 
non trivial solutions of the equation of motion while at the quantum
level it must be retained in order to preserve Weyl invariance.

\subsection{The classical equations of motion}
It is almost immediate to find the general solution of the e.o.m.
associated to the previous action in Minkowski spacetime, i.e. 
when $G_{\mu\nu}=\eta_{\mu\nu}$  as we set in the rest of the paper.
In particular the $A_\alpha$ e.o.m. reads 
\begin{equation}
 \partial_\alpha 
\left( F^{\alpha\beta}  / \sqrt{ -\det (\hat G +\kappa F) }
\right)=0
\end{equation}
which implies
\begin{equation}
F_{0 1}  / \sqrt{ -\det (\hat G +\kappa F) }
= c_0
\end{equation}
with $c_0$ an arbitrary constant which can be identified with
$(\Pi^1+ \T \theta)/(\T \kappa^2)$ where $\Pi^1$ is the $A_1$ conjugate
momentum given in eq. (\ref{Pi1}).
It then follows that we can compute $F_{01}$ as
\begin{equation}
 F_{0 1}^2 = - \frac{ c_0^2}{1+ \kappa^2 c_0^2 } \det \hat G
\end{equation}
Using this result into the action, even if $A_\alpha$ is dynamical and
therefore it is not completely correct, implies that we can rewrite the
action up to topological terms as
\begin{equation}
S= 
-\frac{\T}{ \sqrt{1+ \kappa^2 c_0^2} }  \int d^2 \xi 
\sqrt{ -\det (\hat G_{\alpha\beta} )}
\end{equation}
that is we find a usual bosonic string with a rescaled tension.
In doing so we are nevertheless forgetting the fact that $A_\alpha$ is
dynamical, even if it has only one d.o.f. corresponding to the Wilson
line $w=\oint d \sigma A_1$.
The consequences of this sloppy treatment is that we are, for example, missing 
the relative amplitudes normalizations
%and other relative information 
of the different sectors of the theory
which are characterized by the $\Pi^1$ eigenvalues
and that we can be induced to believe that $F_{01}$ is a constant
operator at the quantum level which is not as eq. (\ref{true-F01}) shows.

In the orthonormal gauge $\dot X^2+ X^{'2}= \dot X \cdot X'=0$
supplemented by the temporal gauge $A_0=0$ the general solution of
e.o.m reads
\begin{align}
X^\mu(\sigma,\tau)
&=x^\mu + \frac{1}{\pi T} p^\mu \tau
+ \frac{1}{2 \sqrt{\pi T}} \sum_{n\ne 0}
\frac{sgn(n)}{ \sqrt{|n|} }
\left(
a^\mu_n e^{-i \frac{2\pi n}{\cl}(\tau+\sigma)}
+ \tilde a^\mu_n e^{-i \frac{2\pi n}{\cl}(\tau-\sigma)}
\right)
\\
A_1(\sigma,\tau)
&=a_1
+ \frac{c_0}{ \sqrt{1+ \kappa^2 c_0^2}}
\int d \tau~ X^{'2}
\end{align}
where we have used the residual gauge invariance to fix the $\tau$
independent integration function appearing in $A_1$ to a constant
$a_1$ and set $T=\sqrt{\T^2+\frac{1}{\kappa^2}(\Pi^1+\T \theta)^2}$.

\subsection{Constraints algebra}
It is immediate to pass to the Hamiltonian formalism and compute the momenta
\begin{align}
\cP_\mu&=
\T
\frac{\dot X_\mu ~X^{'2} - X'_\mu~\dot X \cdot X'}
{ \sqrt{ -\dot X^2 ~X^{'2} + (\dot X \cdot X')^2 - \kappa^2 F^2_{01}  } } 
\\
\Pi^1 &=
\T \kappa^2
\frac{F_{01}}
{ \sqrt{ -\dot X^2 ~X^{'2} + (\dot X \cdot X')^2 - \kappa^2
    F^2_{01}  } } 
-\T \theta
\label{Pi1}
\end{align}
and the Hamiltonian
\begin{align}
H&=
\int_0^\cl d\sigma  (-\Pi^1 A'_0)
\end{align}
so that the primary and secondary constraints can be written as
\begin{align}
L_\pm
&=
(\frac{\cP}{T} \pm T X')^2
\\
\Pi^0&=0
\\
\Pi^{1}{}'&=0
\end{align}
where we defined the  a priori momentum dependent tension
\begin{equation}
T^2(\sigma) =\T^2 + \frac{1}{\kappa^2}(\Pi^{1}(\sigma))^2
=\T^2 S^2(\sigma)
\end{equation}
and the scaling $S(\sigma)$ which plays a major role later, see
eq. (\ref{Scaling}). 
These constraints satisfy an algebra whose non vanishing elements are
the Virasoro algebra  
\begin{equation}
\{ L_\pm(\sigma_1),L_\pm(\sigma_2)\}
=\pm 4 \left(L_\pm(\sigma_1)+L_\pm(\sigma_2) \right) 
~\partial_{\sigma_1} \delta(\sigma_1-\sigma_2)
\end{equation}
and a further constraint
\begin{equation}
\{ L_+(\sigma_1),L_-(\sigma_2)\}
= 4 
(\frac{\cP}{T} + T X') \cdot
(\frac{\cP}{T} - T X')
~\partial_{\sigma_1}(T^{-1})
~\delta(\sigma_1-\sigma_2)
\end{equation}
which involves a structure function which makes the computation of the
BRST charge more involved than usual \cite{Henneaux:1985kr}.
If we could impose at the constraints algebra level the constraint
$\Pi^{1}{}'=0$ we would recover the usual algebra but this is not correct.
Therefore in order to avoid the issues involved in computing the BRST
charge we have chosen to
quantize the theory on the light cone.

\subsection{Light cone quantization}
We can proceed in fixing all the gauge invariances in the light
cone gauge.
In particular we fix Weyl invariance by setting
\begin{equation}
\det \gamma=-1
\end{equation}
Notice that we could have used the condition $\det g =-1$ as well 
and we would have obtained the same results but the chosen choice
makes computations easier.
After this first step we fix almost all worldsheet diffeomorphisms by
\begin{equation}
X^+(\sigma,\tau)=\chi ~(\tau-\tau_0)
\label{gfX+}
\end{equation}
with $\chi=\pm 1$. In this section $\chi=1$  but the choice $\chi=-1$ 
is needed when discussing the interactions, see figure \ref{fig:3interaction}.
We are then left with residual diffeomorphisms given by $\tau=\tau',~
\sigma=\sigma(\sigma',\tau')$  which can be fixed by
\begin{equation}
\gamma_{11}(\sigma,\tau)=g_{11}(\sigma,\tau)= \hat \gamma_{11}(\tau)
\end{equation}
This can be done because   
$ \frac{\gamma_{11}(\sigma,\tau)}{\sqrt{ -\det \gamma} } d \sigma$ 
is invariant under these residual diffeomorphisms and can be used to
introduce a new worldsheet spacial coordinate as
$ \frac{N(\tau)}{\cl} d\zeta
=
 \frac{\gamma_{11}(\sigma,\tau)}{\sqrt{ -\det \gamma} } d \sigma
$.
After this step there are still some residual diffeomorphisms given by
$\tau=\tau',~
\sigma=\sigma'+\sigma_0(\tau')$ which could be used to fix
$\gamma_{01}(\sigma=0,\tau)=0$ but we prefer not to fix it and get the
constraint associated with its e.o.m. when we compute the Hamiltonian.

We can now consider the gauge fixing of the $U(1)$ worldsheet
symmetry. 
Since we are working with a worldsheet with a topology of a cylinder 
we can only set
\begin{equation}
A_1(\sigma,\tau)= a_1(\tau),~~~~
a_1(\tau)\equiv a_1(\tau) + \frac{2\pi n}{\cl}~~n\in\Z
\end{equation}
but we cannot set $a_1=0$ since $\exp \left( i \oint d\sigma ~
A_1\right)$ is gauge invariant.
As it happened with diffeomorphisms 
we are left with a residual gauge symmetry with parameter
$\epsilon=\epsilon(\tau)$ which can be fixed by
\footnote{
In the following for a generic field $Q$ 
$Q_{nzm}$ is the projection of the field $Q$ onto 
the periodic part explicitly $\sigma$ dependent, i.e.
given $Q_{zm}= \frac{1}{\cl} \int_0^\cl d\sigma ~Q$ we set
$Q_{nzm}=Q-Q_{zm}$.
}
\begin{equation}
\oint d\sigma~ A_0=0 \leftrightarrow A_0(\sigma,\tau)=A_{0, nzm}(\sigma,\tau)
\end{equation}
After the gauge fixing the action becomes
\begin{align}
S
&=-\frac{\T}{2}
\int d\tau \Bigl\{
2 \chi \hat g_{11}(\tau) \int_0^\cl d \sigma \dot X^-
+ \dot a_1  \int_0^\cl d \sigma ( 2 \kappa f + 2 \theta)
\nonumber\\
&+ \int_0^\cl d \sigma
\Bigl[
 - ( 2 \kappa f + 2 \theta) A'_{0 nzm}
- \hat g_{11}(\tau) \dot{ X_i}^2
+ \frac{1 +f^2 -g_{01}^2}{  \hat g_{11}} { X_i}'^2
\nonumber\\
&+
2 g_{01} ( - \chi X'^- + \dot X_i X'_i )
+ \lambda \sqrt{1+f^2}
\Bigr]
\Bigr\}
\label{S2}
\end{align}
To further simplify it we can solve for the non dynamical fields. 
From the $A_{0 nzm}$ e.o.m. we get
\begin{equation}
f(\sigma,\tau) = \hat f(\tau)
.
\end{equation}
The variation of the conjugate variable $f_{nzm}(\sigma,\tau)$ yields simply
\begin{equation}
A_{0 nzm}(\sigma,\tau)
= \frac{\hat f(\tau)}{\kappa~  \hat g_{11}(\tau)}
\int d\sigma~ (X_i')^2
\end{equation}
because the $\sigma$ integration of the e.o.m. has no ambiguous
integration constant since $A_{0 nzm}$ has no zero modes.
Notice however that $\hat f(\tau)$ is dynamical and it is proportional
to the momentum conjugate to $a_1$ as eq. (\ref{p}) shows.

The next non dynamical variable we consider is
$X^-_{nzm}(\sigma,\tau)$ whose e.o.m. gives
\begin{equation}
g_{01}(\sigma,\tau)= \hat g_{01}(\tau)
.
\end{equation}
As before the integration of 
e.o.m. of the conjugate variable $g_{01~nzm}(\sigma,\tau)$
yields
\begin{equation}
\chi X^-_{nzm}(\sigma,\tau)
=\int d\sigma~  \left[
\dot X_i X_i'
-\frac{\hat g_{01}(\tau)}{\kappa~  \hat g_{11}(\tau)} (X'_i)^2
\right]
.
\end{equation}
Inserting into the action (\ref{S2}) the previous results we get
\begin{align}
S
&=
\int d\tau \Bigl\{
-\T \cl \chi \hat g_{11}(\tau)  \dot x^-
-\T \cl ( \kappa \hat f + \theta) \dot a_1
- \frac{\T\cl}{2} \lambda \sqrt{1+\hat f^2}
\nonumber\\
&+ \int_0^\cl d \sigma
\Bigl[
\frac{\T\cl}{2} \hat g_{11}(\tau) \dot{ X_i}^2
-\frac{\T\cl}{2} \frac{1 +\hat f^2 -\hat g_{01}^2}{  \hat g_{11}} { X_i}'^2
-\T \hat g_{01} \dot X_i X'_i 
\Bigr]
\Bigr\}
\label{S-gauge-fixed}
\end{align}
from which we can read the momenta
\begin{align}
p^+ &= -p_-= \T \cl \chi \hat g_{11}(\tau)
\label{p+}
\\
p &= -\T \cl ( \kappa \hat f(\tau) + \theta)
\label{p}
\\
\cP_i &= \T ( \hat g_{11} \dot X_i - \hat g_{01} X'_i)
\label{cP}
\end{align}
where $p$ is the conjugate momentum to $a_1$.
The Hamiltonian is then easily computed to be
\begin{align}
H
&=
\frac{\cl}{p^+}
\int_0^\cl d\sigma~  \left[
\oh \cP_i^2
+\frac{\T^2}{2} S^2
X^{'2}_i
+\frac{\T\cl}{p^+} \hat g_{0 1} \cP_i X'_i
\right]
+ \lambda \frac{\T\cl }{2} S
\label{H0}
\end{align}
where we have introduced the scaling 
\begin{equation}
S(p_w)=
\left( 1+ \frac{1}{\kappa^2}\left( \frac{p_w}{\T} + \theta\right)^2
\right)^{1/2}
\label{Scaling}
\end{equation}
with $p_w=p/\cl$ as defined in the last of eq.s (\ref{canon-trans}).

We can now compute the e.o.m in either the Lagrangian formalism or the
Hamiltonian one and find that $p^+,p$ are constant,
\begin{align}
\dot x^- &= - \frac{H}{p^+}
\nonumber\\
\dot a_1 &= \frac{\T}{\kappa^2 p^+} \left( \frac{p}{\T\cl}+\theta
\right)
\int_0^\cl d \sigma   X^{'2}_i 
\end{align}
and at the same time the remaining Lagrange multiplier $\hat g_{01}$  implies
the constraint
\begin{equation}
\int_0^\cl d \sigma  \dot X_i X'_i 
=
\int_0^\cl d \sigma  \cP_i X'_i  =0
\end{equation}
while the $X^i$ e.o.m reads as usual
\begin{equation}
{\ddot X}^i - %\left(\frac{\T \cl}{ p^+} S \right)^2 
\omega^2(p)
%X^{''i}&=0
X'{}'{}^i=0
\end{equation}
with 
$
\omega(p)= \frac{\T \cl}{ p^+} S(p)
$.
%\footnote{
It is worth noticing that
using the $a_1$ e.o.m. we can then compute the on shell expression for
the electric field to be the gauge invariant expression
\begin{equation}
F_{0 1}= \frac{ T \cl}{p^+ \kappa^2} \left( \frac{p}{ T \cl} +
\theta\right) { X_i'}^2
.
\label{true-F01}
\end{equation}
It is then easy to realize how $F_{0 1}$ is not diagonalized on the
the mass eigenstates (\ref{mass-eigenstates}) 
and therefore differently from what
asserted in the literature \cite{Kallosh:1997nr} where it
is assumed that $F_{0 1}$ can be diagonalized the closed $D1$ string has
always massless excitations.
%}

In order to compute the commutation relations
between the modes we write the $X^i$ and $\cP_i$ expansions explicitly as
\COMMENTOO{Controllare $|p^+|$}
\begin{align}
X^i
&=
x^i+ \frac{p^i}{p^+} \tau
+i \sqrt{\frac{\cl}{ 4\pi |p^+|~ \omega(p)}}
\sum_{n\ne 0}
\frac{sgn(n)}{\sqrt{|n|}}
\left(
a^i_n e^{-i \frac{2\pi n}{\cl}(\omega(p) \tau+\sigma)}
+ \tilde a^i_n e^{-i \frac{2\pi n}{\cl}(\omega(p) \tau-\sigma)}
\right)
\nonumber\\
\cP_i
&=
\frac{p^i}{\cl}
+\frac{1}{2\cl} \sqrt{ \frac{ 4\pi |p^+|~ \omega(p)}{l} }
\sum_{n\ne 0}
{\sqrt{|n|}}
\left(
a^i_n e^{-i \frac{2\pi n}{\cl}(\omega(p) \tau+\sigma)}
- \tilde a^i_n e^{-i \frac{2\pi n}{\cl}(\omega(p) \tau-\sigma)}
\right)
\end{align}
As a consequence of the presence of $\omega(p)$ 
which does depend on $p$
in the previous
expansions we find that the operators $a^i, \tilde a^i$ 
do not commute with $a_1$.
The proper way to proceed is to perform a canonical transformation 
at $\tau=0$ for simplicity
and
introduce the new fields as\footnote{ The symmetrization of the
  product $\cP X$ in the definition of $w$ is necessary in order to get
  an Hermitian operator.}
\COMMENTOO{Dipendenza da $\tau$?}
\begin{align}
\hat X^i(\sigma) 
&=
%\left( 1+ \frac{1}{\kappa^2}\left( \frac{p_w}{\T} + \theta\right)^2
%\right)^{1/4}
\sqrt{S(p)}~
X^i(\sigma)
\nonumber\\
\hat \cP_i(\sigma) 
&=
%\left( 1+ \frac{1}{\kappa^2}\left( \frac{p_w}{\T} + \theta\right)^2
%\right)^{-1/4}
\frac{1}{\sqrt{S(p)} }~
\cP_i(\sigma)
\nonumber\\
\hat x^- &= \sqrt{S(p)} ~x^-
\nonumber\\
\hat p^+ &= \frac{1}{\sqrt{S(p)} }~ ~p^+
\nonumber\\
w
&=
\cl a_1 +
\frac{\partial}{\partial p_w}
%\left( 1+ \frac{1}{\kappa^2}\left( \frac{p_w}{\T} + \theta\right)^2
%\right)^{1/4}
\ln(\sqrt{S(p)})~
%\nonumber\\
%&
%~\times
\Big[
\int_0^\cl d\sigma \frac{
X^i%(\sigma,\tau) 
\cP_i%(\sigma,\tau)
+
\cP_i%(\sigma,\tau) 
X^i%(\sigma,\tau)
}{2}
-
\frac{x^- p^+ + p^+ x^-}{2}
\Big]
\nonumber\\
p_w
&=
\frac{p}{\cl} 
\label{canon-trans}
\end{align}
with non vanishing  commutation relations
\begin{align}
[w, p_w] &= i
\\
[\hat x^-, \hat p^+] &=-i
\\
[\hat x^i, \hat p_j] &= i \delta^i_j
\\
[a^i_m, a^j_n] &= [\tilde a^i_m, \tilde a^j_n] = \delta_{n,m} \delta^{i j}
\end{align}
The new operator $w$ corresponds physically 
to the Wilson line $\oint d \sigma ~ A_1$
and it is defined modulo $2\pi $ times an integer, i.e. $w\equiv w+2\pi$
because of the big gauge transformations generated by $\epsilon=
\exp\left(i 2\pi n \frac{\sigma}{l} \right)$.

At this stage one could wonder whether to
stick with  $x,p$ or use $\hat x, \hat p$ since both couples have the same
commutation relation and in both cases one could 
define an operator $w$ in a way to commute with either of them by
using $\int_0^\cl d\sigma (X^i_{nzm} \cP_{i~nzm}+  \cP_{i~nzm} X^i_{nzm})$
or as done before $\int_0^\cl d\sigma (X^i \cP_{i}+  \cP_{i} X^i)$.
One of the main results of this paper is that the proper variables are
$\hat x, \hat p$ since only using them we can factorize amplitudes in
the usual way.

\subsection{Fock space and critical dimension}
We proceed as usual to define the vacuum as
\begin{equation}
a^i_n |0\rangle=\tilde a^i_n |0\rangle=
p^i |0\rangle=p^+ |0\rangle= p_w |0\rangle=0,~~~~
n>0
\end{equation}
and then construct the Fock space by
\begin{equation}
|\{N,\tilde N\}, k^i,k^+,n_w\rangle
=
\prod_{i=1}^{D-2}\prod_{n_i=1}^\infty
\frac{ a_{-n_i}^{i~ N_{n_i}} }{ \sqrt{ %\phantom{ \tilde} 
N_{n_i}!} }
\frac{ \tilde a_{-n_i}^{i~ \tilde N_{n_i}} }{ \sqrt{\tilde N_{n_i}!} }
e^{i \hat k_i \hat x^i + i \hat k^+ \hat x^- +i n_w w} |0\rangle
\label{mass-eigenstates}
\end{equation}

Now we have well defined normal ordering prescription we can
compute the Hamiltonian 
\begin{align}
H
%=P^-
= \frac{\sqrt S }{2 \hat p^+} \left[ (\hat p_i)^2
+ 4\pi \T \left( \sum_{n=1}^\infty n (a^i_{-n} a^i_{n}+ \tilde a^i_{-n}
\tilde a^i_{n}) - \frac{D-2}{12} \right) \right]
\end{align}
As usual we have reabsorbed the divergence from the regularized normal
ordering  constant 
$\frac{S}{2 p^+} \cdot 4\pi \T \cl \cdot 2 (D-2) \sum_{n=1}^\infty n
e^{-\epsilon  2 \pi n / (\cl \sqrt{ g_{11}} )} = - \frac{D-2}{
  \epsilon^2} \frac{p^+ \omega}{2 \pi \T} - \frac{D-2}{12} \frac{2 \pi
  \omega}{\cl} +O(\epsilon)$ into a shift
of the two dimensional cosmological constant $\lambda$ in
eq. (\ref{H0}) while the constant part, regularization independent is
the zero point energy.

It would therefore  seem that the spectrum of the theory is sector
dependent but it is not. There are two reasons why we need rescaling
$P^-$. 
The first is that if we can hope to have a Poincar\'e invariant theory
we cannot rescale all translator generators but $P^-$.
The second is that in order to factorize amplitudes with more 
than 4 legs we need to use hatted zero modes 
%both longitudinal and lightcone ones
as discussed in the next section. 
Therefore we need the hatted Hamiltonian
\begin{equation}
\hat P^-\s2= \hat H\s2 = \frac{1}{\sqrt{S}} H
\end{equation}
which commutes with $w$ differently from $H=P^-$ which does not 
commute with $w$, i.e. $[H,w]\ne0$.
As a consequence the mass spectrum in term of the hatted operators
is independent on the sector $p_w$:
\begin{equation}
M^2= 2 \hat P^-\s2 \hat p^+ -(\hat p_i)^2
= 4\pi \T  \left( \sum_{n=1}^\infty n (a^i_{-n} a^i_{n}+ \tilde a^i_{-n}
\tilde a^i_{n}) -2 \right)
\end{equation}

Let us now discuss the Lorentz invariance and the critical dimension
of the theory.
The non dynamical Lorentz generators are exactly the same as in the
usual theory when expressed using hatted operators. 
The dynamical generators break by definition the gauge
condition (\ref{gfX+}) which must be restored by a diffeomorphism.
On the other side $w=\oint d \sigma A_1$ is diffeomorphism invariant
therefore the dynamical generators have the same expressions as in the
usual theory  when written using the hatted operators 
and hence the critical dimension is left unchanged.

\section{Interactions}
We now determine the three strings interaction vertex by generalizing
the overlapping conditions used by Cremmer, Gervais, Kaku and Kikkawa
\cite{overlap} 
to the case where there are the worldsheet metric and a gauge field on
the world sheet. The same vertex, but only for on shell states,
 can be obtained by writing the vertex operators associated to 
 the physical states obtained from DDF operators \cite{Del Giudice:1971fp}  
and then  computing the string amplitudes
% \cite{Sciuto:1969vz} 
as explicitly done in \cite{Ademollo:1974kz}.

\begin{figure}[hbt]
\begin{center}
\def\svgwidth{350px}
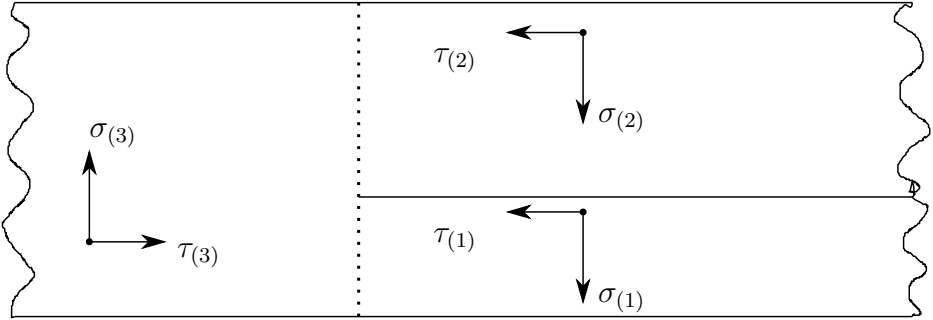
\end{center}
\vskip -0.5cm
\caption{Interaction among three strings with their local worldsheet
  coordinates.} 
\label{fig:3interaction}
\end{figure}

We consider the case where string no. 3  with width $\cl_3$ is incoming
and  splits into outgoing string no. 1 and no. 2 with width $\cl_1$ and
$\cl_2$ respectively. 
The local worldsheet coordinates on the three string are related at
the interaction point $(\sigma_{i},\tau_i)$ as
\begin{align}
\sigma\s3=
\left\{\begin{array}{ccc}
\sigma_{i} \frac{\cl_1 -\sigma\s1}{\cl_1}
& 0\le \sigma\s1 \le \cl_1
& [ 0\le \sigma\s3 \le \sigma_{i 3}]
\\
\sigma_{i} +(\cl_3 -\sigma_{i3}) \frac{\cl_2 -\sigma\s2}{\cl_2}
& 0\le \sigma\s2 \le \cl_2
& [ \sigma_{i 3}\le \sigma\s3 \le \cl\s3]
\end{array}
\right.
\end{align}
and
\begin{equation}
\tau\s3 -\tau_{0(3)}= - (\tau\s1-\tau_{0(1)}) = - (\tau\s2-\tau_{0(2)})
\label{tau-over}
\end{equation}
as shown in figure (\ref{fig:3interaction}).

Let us consider the overlapping condition at $\tau=\tau_i=0$  for
$\det \gamma$, it reads
\begin{equation}
\det \gamma\s3 = 
\left\{\begin{array}{cc}
\det \gamma\s1 (- \frac{d \sigma\s1}{d \sigma\s3})^2
& 0\le \sigma\s1 \le \cl_1
\\
\det \gamma\s2 (- \frac{d \sigma\s2}{d \sigma\s3})^2
& 0\le \sigma\s2 \le \cl_2
\end{array}
\right.
\end{equation}
Since the previous condition must be consistent with the Weyl gauge
fixing we deduce
\begin{equation}
\frac{\sigma_i}{l_1}=\frac{l_3-\sigma_i}{l_2}=1,
\label{Mand-param}
\end{equation}
and with the help of the overlap condition
for the worldsheet metric component $g_{11}$
\begin{equation}
g_{11 (3)} = 
\left\{\begin{array}{cc}
g_{11 (1)}
%\det \gamma\s1 (- \frac{d \sigma\s1}{d \sigma\s3})^2
& 0\le \sigma\s1 \le \cl_1
\\
g_{11 (2)}
%\det \gamma\s2 (- \frac{d \sigma\s2}{d \sigma\s3})^2
& 0\le \sigma\s2 \le \cl_2
\end{array}
\right.
\end{equation}
we obtain Mandelstam parametrization since $p^+ =
\T \cl \chi \hat g_{11}$.

Next we can exam the overlap conditions for the coordinates $X^\mu$
which are then expressed as
\begin{align}
X^i\s3(\sigma\s3,\tau^-_i)
&=  
\left \{ \begin{array} {c c}
X^i\s1(\sigma\s1,\tau^+_i) 
& 0\le \sigma\s1 \le \cl_1
\\
X^i_{(2)}(\sigma\s2,\tau^+_i) 
& 0\le \sigma\s2 \le \cl_2
\end{array} \right.
\nonumber\\
x^-\s3&=x^-\s1=x^-\s2
\end{align}
to which one has to add eq. (\ref{tau-over}) which expresses the $X^+$
overlap after the gauge fixing. It is the $X^+=\chi (\tau-\tau_0)$ overlap 
along with the symmetric choice of worldsheet coordinates pictured in
fig. (\ref{fig:3interaction}) which forces the $X^+$ gauge fixing with
$\chi\s{1,2}=-1$.

It is also worth writing the overlap condition for the worldsheet 1-form current 
$J_\mu={}_2 * \left(\frac{\delta S}{ \delta \partial_\alpha X^\mu} d \xi^\alpha\right)
= \cP_\mu ~ d\sigma+ \dots
$
which can then expressed as
\begin{align}
\nonumber\\
\cP_{i (3)}(\sigma\s3,,\tau^-_i)
&=  
\left \{ \begin{array} {c c}
-\frac{\cl_1}{\sigma_i}\cP_{i (1)}(\sigma\s1,\tau^+_i) 
& 0\le \sigma\s1 \le \cl_1
\\
-\frac{\cl_2}{\cl_3-\sigma_i} \cP_{i (2)}(\sigma\s2,\tau^+_i) 
& 0\le \sigma\s2 \le \cl_2
\end{array} \right.
\nonumber\\
&p^+\s3+p^+\s1+p^+\s2=0
\end{align}
The condition for $p^+$ is also consistent with the overlap condition
for the worldsheet metric component $g_{11}$,
the $X^+$ gauge fixing $\chi\s3=-\chi\s{1,2}=1$,
the constraint (\ref{Mand-param}) on the parametrization and the explicit
expression (\ref{p+}) which relates $p^+$ and $g_{11}$.

To these usual conditions we must add the conditions that follow from
the $A_\alpha d \xi^\alpha$ overlap conditions
\begin{align}
A_{1 (3)}(\sigma\s3,\tau^-_i) ~d \sigma\s3
&=  
\left \{ \begin{array} {c c}
 [ A_{1 (1)}(\sigma\s1,\tau^+_i) +  \partial\epsilon\s1(\sigma\s1)] d \sigma\s1 
& 0\le \sigma\s1 \le \cl_1
\\
{} [ A_{1 (2)}(\sigma\s2,\tau^+_i) + \partial \epsilon\s2(\sigma\s2) ] d \sigma\s2
& 0\le \sigma\s2 \le \cl_2
\end{array} \right.
\nonumber\\
A_{0 (3)}(\sigma\s3,\tau^-_i)
&=  
\left \{ \begin{array} {c c}
 -A_{0 (1)}(\sigma\s1,\tau^+_i) 
& 0\le \sigma\s1 \le \cl_1
\\
-A_{0 (2)}(\sigma\s2,\tau^+_i)
& 0\le \sigma\s2 \le \cl_2
\end{array} \right.
\end{align}
%when we impose the gauge fixing.
The previous conditions are the natural ones when we work in
Hamiltonian formalism where we have only the freedom of performing
$\sigma$ redefinitions.
After the gauge fixing the previous conditions become
\begin{align}
e^{i (\cl_3 a_{1(3)}+\cl_1 a_{1(2)}+\cl_2 a_{1(2)})}&=1
\\
\frac{p\s3}{\cl_3}=\frac{p\s1}{\cl_1}=\frac{p\s2}{\cl_2}&
%p_{w(3)}=p_{w(3)}=p_{w(3)}&
\label{pw-over} 
\end{align}
We use the exponential version for the $\cl a_1$ condition because the Wilson
line $\cl a_1$  is defined modulo $2\pi$ times an integer. 
Finally the condition on $p$ is the only one which is
compatible with the $\cl a_1$ condition and the
commutation relations, moreover it can
be directly derived from the expression (\ref{p}) which relates $p$
and $\hat f$ and the continuity condition for $\hat f$.

We want now proceed in computing the interaction 3-vertex
$|V_{3}\rangle$ by imposing the previous conditions.
Because of the $p_w= p/l$ overlap condition 
\begin{equation}
\left(p_{w (r)} -p_{w (s)} \right)  |V_{3}\rangle=0
~~~r,s=1,2,3
\label{pw-reflection}
\end{equation}
almost all the previous overlap conditions can be written using hatted
operators simply by substituting the
unhatted operators. Explicitly we can write
\begin{align}
(\hat x^-\s r - \hat x^-\s s)  
|V_{3}\rangle   &=0
\label{hatted-x-overlap-eq}
\\
\sum_{r=1}^3 \hat p^+\s r
|V_{3}\rangle &= 0
\label{hatted-p-overlap-eq}
\\
\left(
\hat X^i\s3(\sigma\s3) 
- \theta_1 \hat X^i\s1(\sigma\s1)
- \theta_2 \hat X^i\s2(\sigma\s2) \right)
|V_{3}\rangle &= 0
\label{hatted-X-overlap-eq}
\\
\left(
\hat \cP^i\s3(\sigma\s3) 
+ \theta_1 \hat \cP^i\s1(\sigma\s1)
+ \theta_2 \hat \cP^i\s2(\sigma\s2) \right)
|V_{3}\rangle &= 0
\label{hatted-P-overlap-eq}
\end{align}
where we have defined $\theta_2=\theta(\sigma_i - \sigma\s3)$ 
and similarly for $\theta_3=\theta(\sigma\s3 -\sigma_i )$.
Care must be nevertheless used in rewriting the overlap condition for
$l a_1$ in term of $w$ since $w$ is
obtained from $l a_1$ with a shift by quantities which are not
normal ordered and a function of its momentum $p_w$.
In particular we start writing
\begin{equation}
e^{i l a_1}= e^{i w} \left(\frac{S(p_{w }+1)}{ S(p_{w }) }\right)^{
\frac{i}{2} 
\left[
\int_0^\cl d\sigma \frac{ X^i \cP_i + \cP_i X^i }{2}
-
\frac{x^- p^+ + p^+ x^-}{2}
\right]
}
\label{exp_iw-1st_step}
\end{equation}
then using eq. (\ref{pw-reflection}) and 
the reflection properties for $\hat X\s3$ and $\hat \cP\s3$ in 
$e^{i  \sum l\s r a_{1 (r)}} |V_{3}\rangle=1$ 
we see that all the terms involving $\hat X^i\s{1,2,3}$ and $\hat \cP_{i  (1,2,3)}$
(or $X^i\s{1,2,3}$ and $\cP_{i  (1,2,3)}$)
cancel and we are left with the contributions from the lightcone zero modes
$\hat x^-$ and $\hat p^+$.
Finally with the help of the reflection properties for $\hat x^-$ and 
$\hat p^+$  we get
\begin{align}
%e^{i (w\s3+w\s1+w\s2)}|V_{3}\rangle
e^{i \sum_{r=1}^{N=3} w\s r}|V_{3}\rangle
=&
\left(\frac{S(p_{w (1)})}{ S(p_{w (1)} -1) }\right)^{
 \frac{1}{4}( N -2)|_{N=3}
}
|V_{3}\rangle
\label{w-overlap}
\end{align}
At first sight it seems curious that the only non trivial contribution
comes from the light cone zero modes $\hat x^-$ and $\hat p^+$ and one
would expect that also the transverse zero modes contribute but, as
discussed in appendix \ref{app:alt-derivation}, the regularized
contribution from the transverse non zero modes cancel the one 
from the transverse zero modes.
There is also an intuitive explanation why the only contribution comes
from $\hat x^-, \hat p^+$  and it is related to the fact that
 only these d.o.f. are treated symmetrically in outgoing and incoming strings 
(as overlap eq.s (\ref{hatted-x-overlap-eq}), (\ref{hatted-p-overlap-eq}) 
and eq. (\ref{scalingx-p+}) show)
and we want the relative normalization of the different $p_w$ sectors to
be independent on whether we consider a $1\rightarrow 2$ or
$2\rightarrow 1$ interaction.

All the previous conditions can be satisfied if we write
\begin{equation}
|V_{3}\rangle
=
c_0~
|V_{3}\rangle_{\hat X}
\otimes
\sum_{n_w \in \Z}
S(n_{w })^{\frac{1}{4}(N-2) |_{N=3} }
|p_{w (1)}=p_{w (3)}=p_{w (3)}=n_w\rangle
\label{final-V3}
\end{equation}
where $c_0 S(0)$ is the string coupling constant in the $n_w=0$ sector,
$|V_{3}\rangle_{\hat X}$ is the usual bosonic 3-string vertex
operator written in term of the hatted operators.
Would we not use hatted quantities we could not have written the
3-vertex in a factorized form since the overlapping matrices would
explicitly depend on $p_{w(r)}=n_w$: this is clearly shown by
eq. (\ref{matrix-dep-on-pw}) in appendix.
Therefore if we use hatted operators we can completely separate the
``stringy'' operators from the worldsheet electric variables.
Since interactions take place only among strings with 
fixed and equal $p_{w (r)}=n_w$
the computations of interacting lightcone Hamiltonian (\ref{full-p-}) and
amplitudes factorization work in the same way as
usual in each sector only when we use hatted operators, in particular
we need also to rescale the Hamiltonian to obtain a hatted Hamiltonian
which is also the $P^- \s2 $ generator, is independent on the sector
$n_w$ and commutes with $w$ differently
from the unhatted Hamiltonian.
%
%We see then that in the hatted basis there is a complete factorization
%in the vertex $|V_3\rangle$ therefore the part which involves $a$ is
%the same as usual with the substitution $x \rightarrow \hat x$.
\subsection{The effective action}
\begin{figure}[hbt]
\begin{center}
\def\svgwidth{350px}
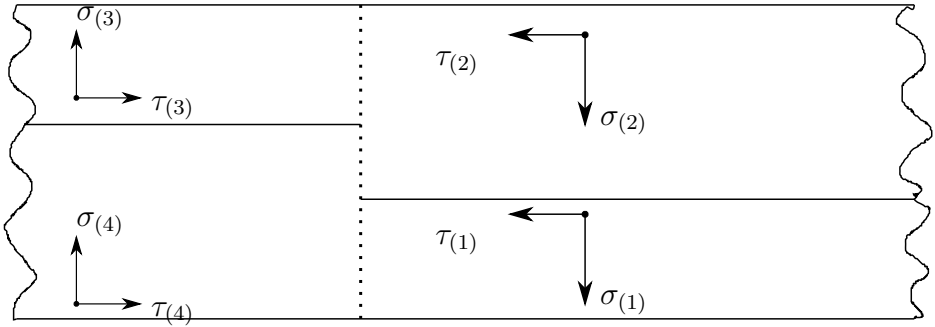
\end{center}
\vskip -0.5cm
\caption{Special configuration of an interaction among four strings
  with their local worldsheet coordinates. The two interactions happen
  at the same time.} 
\label{fig:4interaction}
\end{figure}

The key point is that vertexes and consequently 
amplitudes are completely factorized in a part
which depends on  $\hat X$  times a part which depends on $w$.
Therefore
%the factorization of  amplitudes with more than
%4 legs works in the hatted coordinates.
%Moreover 
the full interacting lightcone Hamiltonian can be 
written as
\begin{equation}
\hat P^- = \hat P^-\s2 + g(p_w) \hat P^-\s3 + g^2(p_w) \hat P^-\s4 + \dots  
\label{full-p-}
\end{equation}
with $\hat P^- \s N$ the $N$ particle contact Hamiltonian which does not
depend on $w$ and $p_w$.
Hence every sector $n_w$ has the same lightcone Hamiltonian and 
the same effective action when written in term of the hatted coordinates.
The only difference is that each sector has a different, rescaled
string coupling constant given by
\begin{equation}
g({n_w})= c_0 ~S(n_w)^{\frac{1}{4}}
=c_0 \left( 1+ \frac{1}{\kappa^2}\left( \frac{n_w}{\T} + \theta\right)^2
\right)^{\frac{1}{8}}
\end{equation}
which can be read from (\ref{final-V3}).
It is also possible to verify directly and easily the dependence on
$n_w$ of the string coupling constant for the cases with $N \ge 4$
interacting strings using the overlapping
conditions for special configurations where all interactions happen at
the same time as pictured in figure (\ref{fig:4interaction}) for the
case $N=4$. In all these case one gets immediately
eq. (\ref{w-overlap}) with the appropriate $N$.

\section{Conclusions}
In this paper we have quantized the bosonic part of the closed $D1$
string action.
We have found that it is viable as a fundamental action and describes
an infinite number of sectors. The sectors are characterized by the
eigenvalue of the momentum $p_w$ of the $U(1)$ Wilson line $w=\oint
A_1 d\sigma$.
Each sector interacts only with strings of the same sector
and has the same spectrum and same interactions as the bosonic string.
The only difference in the interactions is that each sector has a
different string coupling constant.

Actually we have the same interactions as the bosonic string when we
use sector dependent rescaled momenta (and positions).
This rescaling is not at all arbitrary but it is dictated by the
factorization of multipoint string amplitudes.
It follows that all sectors have the same effective action but  
string amplitudes in different sectors are not equal when written
using the natural momenta which are derived from the action
but become equal when kinematical variables are rescaled.
This very same picture continues to hold when the full $D1$ action in
Green-Schwarz formalism is considered \cite{me}.

This suggests a solution of the puzzle which seems to arise when one considers
the action of $SL(2,\Z)$ symmetry on string amplitudes \cite{Garousi0}.
In fact if one acts with a $SL(2,\Z)$ transformation on a string
amplitude and takes the resulting amplitudes as they stand it seems
that for any element of the $SL(2,\Z)$ symmetry we have a different
spectrum.
The solution suggested from this paper is that the kinematical
factors  of amplitudes
obtained by the action of a $SL(2,\Z)$ element  should be rescaled in
such a way to have the same poles as the usual amplitudes.

%%%%%%%%%%%%%%%%%%%%%%%%%%%%%%%%%%%%%%%%%%%%%%%%%%%%%%%%%%%%
\appendix

\section{Another way of performing the main computation}
\label{app:alt-derivation}
In order to elucidate why only the lightcone zero modes contribute to
the $w$ overlapping conditions 
we proceed as done by Kato and Kikkawa which followed Goto and
Naka \cite{Goto:1974uq} and we define
\begin{align}
|V_{3(0)}\rangle
&=
e^{i \int_0^{\sigma_i} d \sigma\s3~ \cP\s3(\sigma\s3)\cdot X\s1(\sigma\s1)
+
i \int_{\sigma_i}^{\cl_3} d \sigma\s3~ \cP\s3(\sigma\s3)\cdot
X\s2(\sigma\s2)
}
|V_3\rangle
\label{V30}
\end{align}
Using this new object the overlapping conditions for $X$ and $\cP$ read
\begin{align}
X^i\s3(\sigma\s3, \tau_i) |V_{3(0)}\rangle
=
\cP_{i (1)}(\sigma\s1, \tau_i) |V_{3(0)}\rangle
=
\cP_{i (2)}(\sigma\s2, \tau_i) |V_{3(0)}\rangle
=&0
%\nonumber\\
%(p_{w (3)} -p_{w (1)} ) |V_{3(0)}\rangle
%=
%(p_{w (3)} -p_{w (2)} ) |V_{3(0)}\rangle
%=&0
\end{align}
which imply the very simple solution
\begin{align}
|V_{3(0)}\rangle
&=
\prod_{i=1}^{D-2} \left[
e^{\sum_{n=1}^\infty \left(
a^{ i \dagger }_{n(3)} \tilde a^{i \dagger}_{n(3)}
-a^{ i \dagger }_{n(1)} \tilde a^{i \dagger}_{n(1)}
-a^{ i \dagger }_{n(2)} \tilde a^{i \dagger}_{n(2)}
%-a^i_{-n(2)} \tilde a^i_{-n(2)}
\right)}
\delta(\hat x^i\s3)
|\hat p^i\s1=\hat p^i\s2=\hat p^i\s3=0\rangle
\right]
\nonumber\\
&\otimes \delta(p^+\s3+p^+\s1+p^+\s2) | x^-\s1=x^-\s2=x^-\s3=0\rangle
\nonumber\\
&\otimes
\sum_{n_w\in\Z} c_{n_w} |p_{w (1)} =p_{w (2)} =p_{w (3)}=n_w\rangle 
\label{V3_0}
\end{align}
with arbitrary $c_n$.
If we compute $|V_3\rangle$ from this explicit expression for $|V_{3(0)}\rangle
$ by inverting eq. (\ref{V30}) we realize immediately that if we do not use
hatted zero modes we get overlapping matrices
explicitly dependent on $p_{w(r)}=n_w$ and we cannot separate the
stringy d.o.f. from the electric one,
for example using the usual notation we would get
\begin{align}
\int_0^{\sigma_i} d \sigma\s3~ 
\hat \cP\s3(\sigma\s3)\cdot \hat X\s1(\sigma\s1)
\supset
\frac{\sigma_i}{2 \cl_3} \sqrt{4\pi \T S(p_{w (3)}) }
\sum_{n} \sqrt{|n|} (a^i_{n (3)} + \tilde a^i_{-n (3)}) x^i\s1 A^{(3
  1)}_{n 0} 
\label{matrix-dep-on-pw}
\end{align}

Now eq. (\ref{exp_iw-1st_step}) can be written in an explicit way
using oscillators as
\begin{align}
e^{i w} 
=\left(\frac{S(p_w)}{ S(p_w -1) }\right)^{
\oh\left[ 
\sum_{n=1}^\infty \left( a^i_{n} \tilde a^i_{n} - a^i_{-n}
  \tilde a^i_{-n} \right)
-i \oh \left( \hat x^i \hat p_i + \hat p_i \hat x^i 
-\hat x^- \hat p^+  - \hat p^+ \hat x^-
\right)
\right] 
}
e^{ i \cl a_1}
%|V_{3(0)}\rangle
%
\end{align}
Using the obvious reflection property for $a$ we read from
(\ref{V3_0}) we get
\begin{align}
e^{i (w\s3+w\s1+w\s2)}|V_{3(0)}\rangle
=&
\left(\frac{S(p_{w (1)})}{ S(p_{w (1)} -1) }\right)^{
\oh\left[ 
(D-2)\sum_{n=1}^\infty 1
+\frac{D-2}{2}
\right]
}
\nonumber\\
&
\prod_{r=2}^3
\left(\frac{S(p_{w (r)})}{ S(p_{w (r)} -1) }\right)^{
\oh\left[ 
-(D-2)\sum_{n=1}^\infty 1
-\frac{D-2}{2}
\right]
}
\nonumber\\
&
\prod_{r=1}^3
\left(\frac{S(p_{w (r)})}{ S(p_{w (r)} -1) }\right)^{
i  \frac{1}{4} \left( 
\hat x^-\s r \hat p^+\s r  + \hat p^+\s r \hat x^-\s r
\right)
}
|V_{3(0)}\rangle
\end{align}
The lightcone zero modes can be treated using
\begin{align}
e^{i \alpha \sum_{r=1}^N \left( 
\hat x^-\s r \hat p^+\s r  + \hat p^+\s r \hat x^-\s r
\right)} 
\delta(\sum_{r=1}^N p^+\s r) | x^-=0\rangle
=e^{( N -2) \alpha }
\delta(\sum_{r=1}^N p^+\s r) | x^-=0\rangle
\label{scalingx-p+}
\end{align}
while regularizing $\sum_{n=1}^\infty 1$ either with $\zeta$ function or as
$\sum_{n=1}^\infty e^{-\epsilon  2 \pi n / (\cl \sqrt{ g_{11}} )}$  as
we did before we get that the finite part is $-\oh$ hence
\begin{align}
e^{i (w\s3+w\s1+w\s2)}|V_{3(0)}\rangle
=&
\left(\frac{S(p_{w (3)})}{ S(p_{w (3)} -1) }\right)^{
 \frac{1}{4}( N -2)|_{N=3}
}
|V_{3(0)}\rangle
\end{align}
as we derived in the main text.

%%%%%%%%%%%%%%%%%%%%%%%%%%%%%%%%%%%%%%%%%%%%%%%%%%%%%%%%%%%%

\end{document}